\documentclass[]{mandm}

\usepackage{array}
\usepackage[utf8]{inputenc}

\usepackage{graphicx}
\usepackage{hyperref}
\usepackage{adjustbox}
\usepackage{subcaption}
\usepackage{float}
\usepackage{placeins}
\usepackage{stfloats}

\usepackage{bm}% bold math
\graphicspath{{./figures/}{./}}
\usepackage{upgreek}
\usepackage{amsmath}

\usepackage{balance}
\usepackage{todonotes}

\jourvolume{XX}
\jourissue{Y}
\jourpubyear{20ZZ}

\begin{document}

\sloppy

\title[Parallel mode differential phase contrast I]{Parallel mode differential phase contrast in transmission electron microscopy, I: Theory and analysis}

%Remove original authors and replace with other stuff
\author[G. W. Paterson et al.]{G. W. Paterson,$^1$
  G. M. Macauley,$^{1,2}$,
  S. McVitie,$^1$
  and Y. Togawa$^{3,1}$}

\affiliation{%
$^1$SUPA, School of Physics and Astronomy, University of Glasgow, Glasgow G12 8QQ, UK.\\
$^2$Current addresses: Laboratory for Mesoscopic Systems, Department of Materials, ETH Zurich, 8093 Zurich, Switzerland; Laboratory for Multiscale Materials Experiments, Paul Scherrer Institute, 5232 Villigen, Switzerland.\\
$^3$ Department of Physics and Electronics, Osaka Prefecture University, Sakai, Osaka, 599-8531, Japan.\\
Corresponding Author: G. W. Paterson \email{Dr.Gary.Paterson@gmail.com}}
  
\begin{frontmatter}

\maketitle

\begin{abstract}
In Part~I of this diptych, we outline the parallel mode of differential phase contrast (TEM-DPC), which uses real-space distortion of Fresnel images arising from electrostatic or magnetostatic fields to quantify the phase gradient of samples with some degree of structural contrast.
We present an analysis methodology and the associated software tools for the TEM-DPC method and, using them together with numerical simulations, compare the technique to the widely used method of phase recovery based on the transport-of-intensity equation (TIE), thereby highlighting the relative advantages and limitations of each.
The TEM-DPC technique is particularly suitable for \textit{in-situ} studies of samples with significant structural contrast and, as such, complements the TIE method since structural contrast usually hinders the latter, but is an essential feature that enables the former.
In Part~II of this work, we apply the theory and methodology presented to the analysis of experimental data to gain insight into two-dimensional magnetic phase transitions.

\noindent\textbf{Key Words:} transmission electron microscopy, differential phase contrast, Lorentz, Fresnel, image distortion
% \noindent(Received XX Y 20ZZ; revised XX Y 20ZZ; accepted XX Y 20ZZ)
\end{abstract}

\end{frontmatter}

\section{Introduction}
\label{sec:intro}
The sub-micron scale characterisation of electro- and magneto-static fields supported by materials has been critical to fundamental research into materials and to the development of data storage and other novel devices.
Within transmission electron microscopy (TEM), several Lorentz parallel beam imaging modes have been developed over the decades and many are in regular use today, including holography~\citep{Gabor_1949_holography, Fukuhara_holography_1983_PhysRevB.27.1839}, small angle electron scattering~\citep{Goringe_1967_saes, Togawa_2013_SAES}, Foucault~\citep{Maron_1948_focault, Nakajima_2016_foucault}, and Fresnel~\citep{Cohen_1967_lorentz, Chapman_1984_mag_methods}.
Of these, Fresnel imaging is perhaps the most common and readily accessible technique, as real-space images containing information on the directional components of the fields may be obtained simply by defocusing the main imaging lens of a standard TEM.

From a wave-optical perspective, the intensity contrast in Fresnel images arises due to the phase change of the electron beam from the electric scalar potential or magnetic vector potential associated with the sample~\citep{AB_PhysRev_1959}.
The change in phase causes the transmitted beam to locally converge or diverge, and this modulates the image intensity obtained at a defocus.
At low defocus, quantitative reconstruction of the phase change is possible through solving the transport-of-intensity equation (TIE)~\citep{Teague_1983_greens_phase} using multiple images recorded at different defocus values~\citep{Bajt_2000_TIE, DeGraef_JAP_2001_quant_TIE}.

Classically, the diverging or converging beam is interpreted as Lorentz deflection of the charged particle.
This was understood in some of the earliest work on quantifying the magnetisation distribution of samples from their induction in a TEM~\citep{Fuller1960_jap_deflection}, but direct use of the real-space deflection information has been rare in the decades since then.
Indeed, only a small number of reports of its use have been made, and these used a modified TEM, where a patterned film or shadow mask was incorporated to allow stray magnetic~\citep{Wade_1976_ToM, Suzuki_1997, Suzuki_ToM_2000, Shimakura_JMSJ_2003} and electric~\citep{Sasaki2010_JEM_apertures} fields to be mapped in vacuum.
When used in tomographic applications, this technique has been labeled `projected electron magnetic tomography' (PEMT)~\citep{Shimakura_JMSJ_2003}.

In almost all samples, some amount of structural contrast exists, from the sample itself or from a support film, and deflection of the beam transmitted through the sample contains information on the phase.
In this work, we investigate this signal and develop a methodology to use it for phase recovery in an unmodified TEM.
The signal produced from this process is proportional to the first derivative of the phase, so we refer to it as TEM-differential phase contrast (TEM-DPC), in order to differentiate it from the related scanning TEM (STEM) technique of STEM-DPC~\citep{Dekkers1974_stem_dpc, Chapman1978_UM_dpc, chapman_1990_mdpc}.

Through numerical simulations we explore the limitations and advantages of the TEM-DPC technique, which are somewhat different to those of the TIE one.
In common with the TIE method, a reference image must be obtained for the TEM-DPC technique, which is straightforward to do in many \textit{in-situ} experiments.
However, as a consequence of using local image displacements rather than intensity changes, the TEM-DPC method is intrinsically less susceptible to systematic errors related to the microscope optics, including changes in magnification, distortion (such as pin cushion), rotation, and illumination that are common to the TIE method as a result of changing focus.
The methodology presented is applicable to a wide range of phase-imaging experiments where the deflection angles are small and a suitable reference image can be obtained.
It complements the TIE method by being particularly suitable for use with images from samples with weak phase contrast compared to structural contrast, a regime to which the TIE method is potentially less well suited.

In Part II of this work~\citep{tem_dpc_part2}, we apply the TEM-DPC technique to experimental cryo-TEM Fresnel images of the magnetic phase transition of a thin lamella of K$_2$CuF$_4$, a material of interest for its quasi \mbox{2-D} ferromagnetism at low temperatures~\citep{Hirakawa_JAP_1982, Togawa2021_JPSJ_transition}.

\section{Phase Induced Image Distortion}
\label{sec:theory}
The theory of electron optics and its application to the study of electromagnetic fields is well developed~\citep{DEGRAEF200127, Beleggia_phil_mag_2003_mag_sim_p1, Beleggia_phil_mag_2003_mag_sim_p2, Zweck_2016_em_review}.
In the following, we cover the parts most relevant to the electron phase change on encountering electromagnetic potentials.

\begin{figure}
  \centering
      \includegraphics[width=6.5cm]{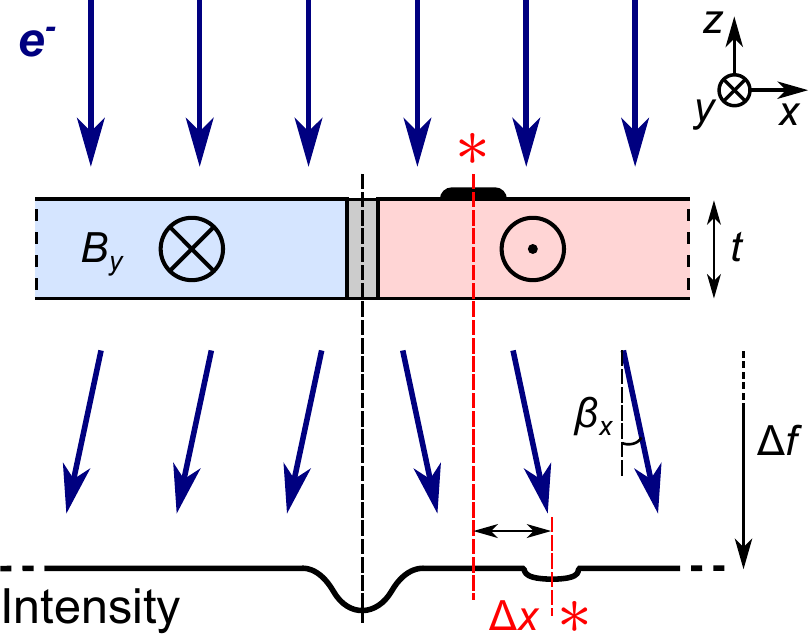}
      \caption{One dimensional schematic of Fresnel contrast from imaging with electrons ($e^-$) a sample of uniform thickness, $t$, and with two regions of antiparallel magnetisation giving rise to induction $\pm B_y$ along the $y$-axis.
      Classically, the electrons are deflected by an angle $\beta_x$ due to the Lorentz force as indicated by the dark blue arrows, creating a reduction in intensity at the projected position of the wall at a defocus $\Delta f$ (underfocus is depicted) due to the diverging beam.
      The apparent position of a localised absorbing area is displaced in the image plane by a distance $\Delta x$ due to the angle through which the beam is deviated (see Eq.~(\ref{eqn:integrated_induction_distortion})). These positions are marked by $\ast$'s.}
      \label{fig:fresnel_schematic}
\end{figure}

For a continuous and slowly varying phase object in the \textit{x-y} plane, a parallel electron beam of wavelength $\lambda$ and which travels along the negative $z$-axis will be deflected by an angle $\bm{\beta}_\perp = (\beta_x, \beta_y)$ which is well approximated by
\begin{equation}
  \bm{\beta}_\perp = -\frac{\lambda}{2\pi} \nabla_{\perp} \phi,
  \label{eqn:beta_phasegrad}
\end{equation}
where $\nabla_{\perp}$ is the gradient operator in the \textit{x-y} plane, $\phi$ is the phase of the exit wave, and $\bm{\beta}_\perp$ is the component perpendicular the $z$-axis, with a sign indicating that of the \textit{x-y} component.
This configuration is depicted in Fig.~\ref{fig:fresnel_schematic}, which is discussed in detail later.
In Fresnel imaging, this deflection creates an apparent change in position in the image plane, $\bm{\Delta r}_\perp = (\Delta r_x, \Delta r_y)$, that is proportional to the defocus, $\Delta f$:
\begin{equation}
    \bm{\Delta r}_\perp = \bm{\beta}_\perp \Delta f,
\end{equation}
where positive defocus corresponds to the underfocus condition (weakened lens excitation).
Together, these two equations relate the phase gradient to the lateral deflection:
\begin{equation}
  \nabla_\perp \phi = -\frac{2\pi}{\lambda} \frac{\bm{\Delta r}_\perp}{\Delta f},
  \label{eqn:phasegrad_displacement}
\end{equation}
which may be solved for the phase using standard Fourier methods or by other means, to within some constant offset which can often be ignored.

The phase change induced in the transmitted beam by the sample is composed of electric ($\phi_e$) and magnetic contributions ($\phi_m$)~\citep{AB_PhysRev_1959}:
\begin{align}
  \phi(x, y) & = \phi_e + \phi_m \\
                & = \frac{\pi}{\lambda E} \int_l V(x, y, z) dz - \frac{\pi}{\Phi_0} \int_l A_z(x, y, z) dz
\end{align}
where $l$ is the electron trajectory, $E$ is the total beam energy, $V$ and $\bm{A}$ are the electrostatic scalar potential and the magnetic vector potential, respectively, and $\Phi_0 = h / 2e$ is the flux quantum, in which $e$ is the electron charge and $h$ is Planck's constant.
Encoded in the spatial dependence of the potentials $V$ and $\bm{A}$, and so the resulting phase contributions, is the geometry of the sample.

The electrostatic component can be expressed in terms of the mean inner potential, $V_0$:
\begin{equation}
    \phi_e = \pi \frac{V_0 t}{\lambda E},
    \label{eqn:e_phase_mip}
\end{equation}
where $t$ is the sample thickness.
The potential includes contributions from the atomic potential, and so is material specific, and also from electrostatics such as polarisation and fixed charge.
The factor $\pi / \lambda E$ is known as the interaction constant (equal to 7.29~mradV$^{-1}$nm$^{-1}$ at 200~kV).
Combining Eqs.~(\ref{eqn:phasegrad_displacement}) and (\ref{eqn:e_phase_mip}) shows how the projected potential can be obtained from the image displacement:
\begin{equation}
  t \nabla_{\perp} V_o + V_o \nabla_{\perp} t = -2 E\frac{\bm{\Delta r}_\perp}{\Delta f}.
  \label{eqn:e_phase_dpc}
\end{equation}
For uniform potentials or thicknesses, the first or second term on the left in Eq.~(\ref{eqn:e_phase_dpc}) is zero, respectively, and this equation may be used to map the sample thickness or the electrostatic field supported by the sample.

While analytical solutions for the magnetisation component of the phase exist in simple geometries~\citep{Beleggia_phil_mag_2003_mag_sim_p1}, it is generally calculated numerically using the Fourier-space approach~\citep{Mansuripur_JAP_1991, Beleggia2003_UM_fourier_phase, Beleggia_apl_2003_mag_sim}, as we do in this work.
In most thin film magnetic samples prepared for TEM characterisation, the thickness is well controlled and only the magnetic component to the phase change gives rise to contrast.
Furthermore, in most materials research, it is not the phase that is of primary interest, but the magnetisation and induction.
The induction component perpendicular to the electron beam, $\bm{B}_\perp$, is related to the phase gradient by~\citep{Chapman1989_mag_structures_tem}:
\begin{equation}
  \nabla_{\perp} \phi_m = - \frac{\pi}{\Phi_0} \int_l (\bm{B}_\perp \times \bm{n}) dz \approx - \frac{\pi}{\Phi_0} \bm{B}_\perp \times \bm{n} t,
  \label{eqn:integrated_induction}
\end{equation}
where $\bm{n}$ is a unit vector normal to the surface (pointing along $z$), and the approximation is for the case of uniform induction throughout the film and no stray field (equivalently, $\bm{B}_\perp$ can be thought of as the projected induction).
Here and below, the cross product simply reflects the nature of the Lorentz force.
Finally, combining Eqs.~(\ref{eqn:phasegrad_displacement}) and (\ref{eqn:integrated_induction}) allows the induction to be calculated from the image displacement:
\begin{equation}
  \bm{B}_\perp \times \bm{n} = \Phi_0 \frac{2}{\lambda t} \frac{\bm{\Delta r}_\perp}{\Delta f}.
  \label{eqn:integrated_induction_distortion}
\end{equation}

In certain situations, the magnetisation distribution, $\bm{M}_\perp$, may be inferred from the induction.
This case is most easily encountered in situations where there is little or no magnetisation divergence, $\nabla_{\perp} \cdot \bm{M}_\perp$, such as in vortex domains, where the induction and magnetisation are simply related by $\bm{M}_\perp \approx \bm{B}_\perp / \mu_0$. 
However, when there is non-zero divergence of magnetisation, then $\bm{H}$ fields will be generated~\citep{Stephen_JAP_2001} which contribute to the measured induction.
This most often occurs at domain walls~\citep{Benitez2015_NatCommun_DW, PhysRevB.99.224429}, where data must be interpreted with care.

The essence of the physics described above is depicted in the simple case of two domains which are uniformly magnetised along the $y$-axis in the schematic of Figure~\ref{fig:fresnel_schematic}.
The diverging transmitted beam on opposite sides of the domain wall creates a reduction in intensity at the projected position of the domain wall.
The intensity distribution within the projected domains is uniform because all electrons are equally deviated, and the image shift is mostly invisible, as depicted on the left side of the wall.
Importantly, however, non-magnetic contrast such as that produced by an absorbing surface feature marked by an $\ast$ in the figure \emph{is} visibly displaced in the projected image.
Much time and effort is put into making high quality samples where the structural contrast is minimised.
However, this can be a very challenging process and some structural contrast always remains.
This contrast is not ideal for conventional TEM (CTEM) or Fresnel imaging, but it does form a potential signal that may be used to assess the image distortions and thus the phase change imparted on the beam by the sample.

\section{TEM-DPC}
The theory outlined in the previous section has been widely known for decades~\citep{Fuller1960_jap_deflection} and is valid when images are formed by the parallel illumination of the sample.
Conventional Fresnel imaging relies on image intensity, and is approximately proportional to the Laplacian of the phase (for weak-phase objects).
From Eq.~(\ref{eqn:phasegrad_displacement}), it is clear that the image distortion analysis approach outlined here may be regarded as a parallel mode differential phase contrast (TEM-DPC) measurement, where the signal generating the contrast (the distortion) is directly proportional to the first derivative of the phase.
The most commonly used technique that produces a similar signal is STEM-DPC, where reciprocal space shifts in the transmitted beam from a focused probe may be used to quantify $\nabla_{\perp} \phi$~\citep{Chapman1978_UM_dpc}.
The technique described here uses localised real-space image shifts to achieve a similar measurement, as one might expect from the reciprocity of TEM and STEM imaging modes, but with additional constraints.

Compared to the in-focus STEM-DPC technique, TEM-DPC must be performed at some degree of defocus and it thus has the much more limited spatial resolution typical of Fresnel imaging~\citep{McVitie2006_UM_fresnel_resolution}.
Just as reference positions are needed in \mbox{STEM-DPC} to quantify the phase-gradients (and a reference image is needed in TIE analysis), a reference image is required in the \mbox{TEM-DPC} method in order to quantify the image distortion field, $\bm{\Delta r}_\perp$, through which the phase gradients may be calculated.
Often, simplifying assumptions about the reference may be made in STEM-DPC, but this is not possible in TEM-DPC, where the reference \emph{must} be spatially resolved.

Another difference between the TEM-DPC and STEM-DPC techniques regards the uniformity of the phase gradient sampled.
This property has been shown to influence the analysis results in the latter method~\citep{Clark2018_pra_dpc_probe}.
We show in a later section that it also affects the applicability and the results of the former method [see Figure.~\ref{fig:linearity}, discussed later].
However, it is worth highlighting here that the relevant scale for each of these techniques is very different: in TEM-DPC (and also in the TIE method), all points of the sample contribute to the formation of the real-space image, whereas the real-space size of the probe at the sample can be sub-nanometer in STEM-DPC.
In principle, a selected area aperture could be used in TEM-DPC to limit the region sampled to some degree but, as we will show, imaging the complete sample in parallel is generally not a major issue and the unmodified technique can still be a useful one.

\begin{figure}
  \centering
      \includegraphics[width=7.5cm]{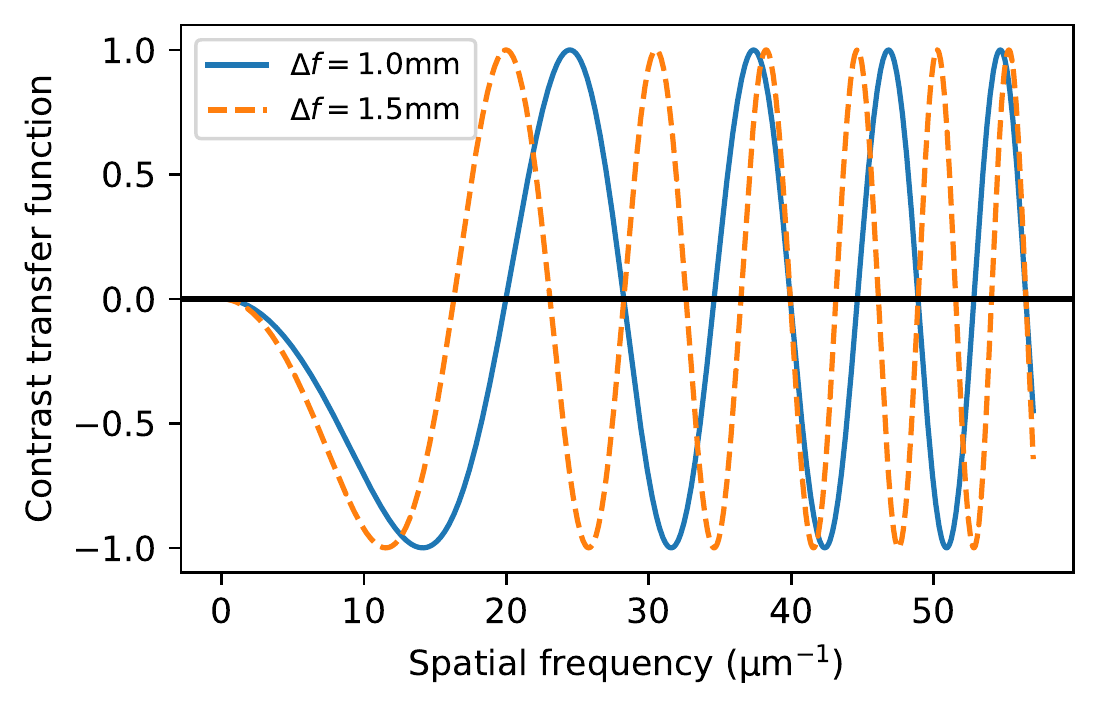}
      \caption{Phase contrast transfer functions at the two different defocus levels indicated in the legend, showing modulation and reversal at higher spatial frequencies.
      The acceleration voltage was 200~kV, and the spherical aberration was taken as zero.
      No envelope functions were applied.}
      \label{fig:ctf}
\end{figure}

For TEM-DPC, the reference image may be one at a different defocus level, an in-focus image, or one obtained at the same defocus but with only the structural phase component present.
The key requirement that must be met in each case is that the same structural features are approximately similar in all images.
This is most easily met in the last of these approaches, as the others may be affected by contrast modification and reversal from the transfer function of the main imaging lens.
Contrast reversal is not only important to consider for the phase signal that we wish to measure, but also for the structural contrast component of the image used for the reference.
The reference contrast tends to consists of higher frequency components and these are more sensitive to changes in defocus.
An example of this effect may be seen in Fig.~\ref{fig:ctf}, which plots the phase contrast transfer function at two defocus values.
The contrast almost completely reverses between the traces at spatial frequencies around 25~$\muup$m$^{-1}$ and 51~$\muup$m$^{-1}$, while being of the same sign at frequencies of 38~$\muup$m$^{-1}$ and 55~$\muup$m$^{-1}$.
There are two main ways of working within this constraint: either use features in the images within a size window that maintains their contrast with defocus or, as we do in this work, maintain focus and modify the polarisation of the sample.
Removal of non-structural phase contrast may be achieved by uniformly polarising the sample or by removing all in-plane polarisation completely by, for example, changing the temperature or applying external magnetic or electric fields.

The TEM-DPC technique is in many ways complementary to the TIE method~\citep{Teague_1983_greens_phase}.
Both techniques are quantitative and subject to transfer function considerations, notably the defocus range.
Whereas TIE uses the change in \emph{intensity} in the linear regime as the beam propagates, TEM-DPC uses the change in \emph{displacement} as the beam propagates.
While the TEM-DPC technique is presented as independent of the TIE one here, features of TIE, such as the changes in intensity from a converging or diverging beam, can in principle be incorporated into the distortion method, which can potentially improve the extraction of the image displacements.
One advantage of the TIE method is that it may be used to recover the information on the phase beyond the sample, thus giving it the ability to map stray fields.
Stray fields may also be measured in the TEM-DPC method, but only when they exist over a suitable substrate material, or where the microscope is modified to incorporate a patterned film or mask~\citep{Suzuki_1997, Suzuki_ToM_2000, Shimakura_JMSJ_2003, Sasaki2010_JEM_apertures}.

Critical to the use of the TEM-DPC technique is the ability to extract the deformation field, $\bm{\Delta r}$, from images.
We discuss various ways of doing this and present our approach in the next section.

\section{Image Distortion Analysis}
\label{sec:align}
Rigid image registration is a standard method used in data processing and encompasses several types of transforms between coordinates with different constraints that preserve some aspect of geometry.
In microscopy, translation transforms are most commonly used to align images with in-plane displacements between them, which often arise due to slow stage drift.
Other transforms with additional degrees of freedom, such as similarity, affine or projective, may be used to correct for additional drifts or changes in view, but are less commonly needed in the parallel imaging modes.
To map image distortion, non-rigid image registration based on local image contrast must be used.
Here, there are generally no geometrical constraints on the transform, though they can be included.
In order to be able to isolate the effects of deformation and drifts of the sample or image position, it is best to rigidly align all data before performing non-rigid registration.
We forgo this initial step for the simulated data in this work.

\begin{figure*}[hbt!]
  \centering
      \includegraphics[width=14.5cm]{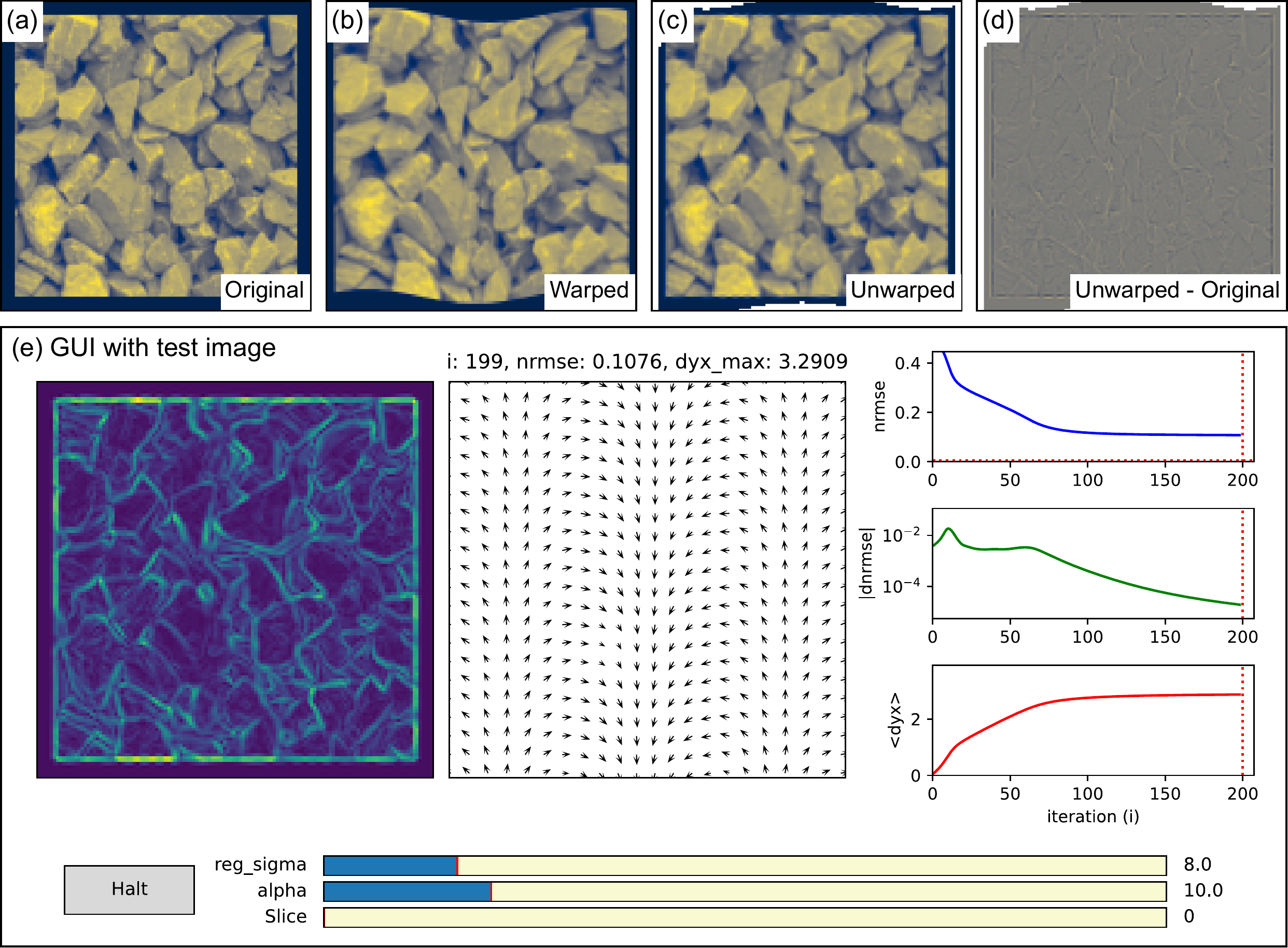}
      \caption{Example of non-rigid alignment using the \texttt{AlignNR} class of the \texttt{fpd} Python library~\citep{fpd} with the `gravel' test image from scikit-image~\citep{scikit-image}.
      (a)-(d) Input and output images: (a) the original image, (b) the warped image (see text for details), (c) the unwarped image after alignment, and (d) the difference between the unwarped and original images; all plotted on a common intensity range, with the range centred at zero in the difference image of (d).
      (e) GUI of the class allowing real-time control and feedback of the image alignment, built using the Matplotlib library~\citep{matplotlib}.
      The plots include the image (left), the distortion field (middle) and extracted metrics (right).
      From top to bottom, the metrics are the normalised root mean square error (nrmse), its change between iterations (on a log scale), and the scale of the vector field extracted (taken as the maximum value).
      The scroll bars allow the regularisation strength (\texttt{reg\_sigma}), the convergence rate (\texttt{alpha}), and the displayed image slice (\texttt{Slice}) to be varied.}
      \label{fig:alignment}
\end{figure*}

Non-rigid image registration has a long history of application in a number of areas.
In structural analysis, digital image correlation (DIC) was developed as a non-contact technique to extract mechanical strain fields in macroscale objects~\citep{Pan_2009_DIC_review, Pan_2018_DIC_dev} and, more recently, applied to higher spatial resolution imaging techniques such as scanning electron microscopy~\citep{Lagattu2006_MC_DIC_sem, vanderesse_2013_DIC}.
Medical imaging is another mature field where optimised non-rigid registration algorithms and libraries~\citep{Klein2010_elastix, Marstal_2016_SimpleElastix} have been developed to enable correlative analysis of images from multiple sources, such as x-ray and magnetic resonance imaging.
In the STEM community, perhaps the most widely known use of non-rigid alignment is in the processing of high spatially resolved images, where multiple images are acquired in quick succession in order to overcome extremely small but unavoidable environmental instabilities~\citep{Jones2015_smart_align}.
The local displacements in such image stacks are typically within a few pixels, where an iterative gradient descent approach based upon the accelerated `demons' method~\citep{Kroon2009_demon_reg} has been shown to work well.

The requirements for the TEM-DPC analysis method outlined in this work are arguably less severe than most others. 
This is due to use of a parallel imaging mode, and because the distortions from the phase profiles typically seen are smooth, especially for magnetic samples where flux lines must be continuous.
Thus, we adopt a similar approach to that in \cite{Jones2015_smart_align}, implemented in the open source Python \texttt{fpd} library~\citep{fpd}.
This allows us full control over the data processing and, in particular, the method used to regularise updates to the distortion field at each iteration of the algorithm.
Our implementation allows for user-provided regularisations to be used.
However, we have found that a simple Gaussian regularisation kernel has the advantage of being far quicker to run than more computationally expensive filters, and it also allows direct control over the spatial resolution through the kernel standard deviation, $\sigma$.
We use this approach for all data in this work.

Figure~\ref{fig:alignment} shows the results of the method used on test data [top row], and the GUI developed to perform the alignment of data [bottom row].
The original image used in the example is shown in Fig.~\ref{fig:alignment}(a), which serves as a reference in the distortion analysis.
Fig.~\ref{fig:alignment}(b) shows this image warped with a distortion field comprised by a superposition of a transverse displacement wave and a compression wave, both traveling along the $x$-axis.
This distortion field was chosen arbitrarily for this demonstration and gives rise to distortion in all directions.
We note that the form of non-rigid alignment used here can also align images between which exist small rotations~\citep{Jones2015_smart_align}, which is a useful property for the analysis of experimental data.  
The distortion field recovered from the alignment procedure is shown as an arrow plot in the centre of Fig.~\ref{fig:alignment}(e).
The `unwarped' image in Fig.~\ref{fig:alignment}(c) is formed by distorting the warped image with the displacement field extracted in the analysis.
The difference between the unwarped and the original images is shown in Fig.~\ref{fig:alignment}(d), where the low contrast confirms that the distortion field has been extracted accurately.

To enhance sensitivity to the high frequency structural contrast and suppress the influence of the low frequency non-structural phase contrast that is our ultimate aim to assess, we use a derivative of a Gaussian filter on the images before performing the non-rigid alignment on the derivative magnitudes.
To a first order, the width of the Gaussian is optimised by choosing a value similar to the edge-width of the non-structural features, although larger values may be appropriate for larger image displacements.
The use of filtering also has the advantage of increasing the convergence rate of the alignment procedure.

Figure~\ref{fig:alignment}(e) shows a screenshot of the GUI tool for image alignment at the end of the analysis.
The plot on the left shows the derivative of the image being aligned [Fig.~\ref{fig:alignment}(b)], at the end of the alignment process.
The arrow plot in the centre shows the final distortion vector field extracted from the data, along with metrics for the alignment, while the plots on the right show the history of the metrics (see caption for details).
All these data are updated in real-time as the alignment proceeds, allowing progress to be tracked, while the scroll bars on the bottom allow the regularisation strength, convergence rate and, when multiple images are being aligned, the slice of dataset shown to be altered.
This real-time control and feedback is critical to allow optimal alignment.
When a suitable regularisation parameter is not known, it is best to start with a large value and then gradually reduce it in order to improve the spatial resolution while monitoring the convergence and the noise level in the vector field plot.
Further discussion of these points accompanied by data from experiment is provided in Part~II of this work~\citep{tem_dpc_part2}.

In the test data above, the texture of the gravel was used to align the data.
In TEM imaging, fine structural features almost always remain from the sample preparation or from contamination, even when much care has been taken during sample preparation.
This contrast goes largely unnoticed in good samples, but can be an issue for imaging in some cases.
These otherwise problematic features can often provide the contrast necessary for the image alignment, either with or without appropriate image filtering.
Where this structural contrast is not intrinsically present in a sample, such contrast may be obtained by placing the sample on a substrate with suitable contrast, or it may be added though deposition of extremely thin films of light materials such as carbon on the sample or substrate, or by ion-beam irradiation of the same.
One particularly promising source of structural contrast for high spatial resolution studies are thin amorphous films, which give rise to weak phase signals with white noise characteristics~\citep{Fan_1986_amorphous}.

The physical scale of the reference contrast sets the spatial resolution limit of the alignment (though not necessarily on a 1:1 basis), and thus one might choose to use very small features when one has control over them.
If the reference image is obtained at a similar defocus as the distorted image, then it does not matter if the high frequency components of the reference signal are inverted, but if the structural contrast changes due to the removal or addition of the non-structural phase, then the alignment will be adversely affected.

For some alignment methods, high frequency reference signals will pose no issue.
However, the gradient descent method employed here can become trapped in local minima, so it is best that the contrast lengthscale is similar to that of the image displacements.
When the main reference contrast is of a finer scale than the image displacements, then bandpass filtering the images or using a large $\sigma$ value can often allow for a successful first alignment.
The procedure may then be iterated with successively less filtering (or filtering over a different band) or smaller $\sigma$ values in order to fully optimise the spatial resolution.
Alternatively, other alignment procedures such as DIC-type approaches, B-spline basis representation~\citep{bunwarpj_2006}, or feature matching are likely to prove successful, with different compromises on the spatial resolution.

As in other analyses, in application of the TEM-DPC method, account must be taken of changes in magnification due to imaging at a defocus.
This may be simply done by calibrating the Fresnel images using an in-focus image of the sample.
At very large defocus or low magnifications, additional distortions such as pincushion and barrel distortion from the microscope optics may occur and the apparent magnification will no longer be uniform across the sample.
If the image against which phase-induced distortions are assessed is recorded under the same conditions as the reference, then the influence of the lens distortion on the phase-induced distortion will be greatly reduced.
However, if required, complete removal of distortions is possible by extracting them by imaging a sample with a known reference shape, recording multiple displaced overlapping images~\citep{Kaynig2010_overlap_distort_corr}, or recording a defocus series where the changes between successive images are small, and then applying the cumulative inverse transform to the source images or the result of their analysis.

\section{Comparison With TIE}
\label{sec:simu}
In the transport of intensity approach to phase retrieval~\citep{Teague_1983_greens_phase} the TIE equation may be solved in the paraxial monochromatic wave approximation using the change in image intensity, $I$, at different defocus values, together with the in-focus image intensity, $I_o$~\citep{Paganin_PRL_1998_noninter_phase}:
\begin{equation}
  \phi = - \frac{2\pi}{\lambda} \nabla_\perp^{-2} \left\{ \nabla_\perp \cdot \left[ \frac{1}{I_o} \nabla_\perp \nabla_\perp^{-2} \frac{\partial I}{\partial z} \right] \right\},
  \label{eqn:tie}
\end{equation}
where $\nabla_\perp^{-2}$ is the inverse Laplacian, and the subscript indicates the plane perpendicular to the beam.
Since its early adoption in TEM~\citep{Bajt_2000_TIE, DeGraef_JAP_2001_quant_TIE}, the TIE technique has been regularly used in the study of magnetic samples.

\begin{figure}
  \centering
      \includegraphics[width=7.5cm]{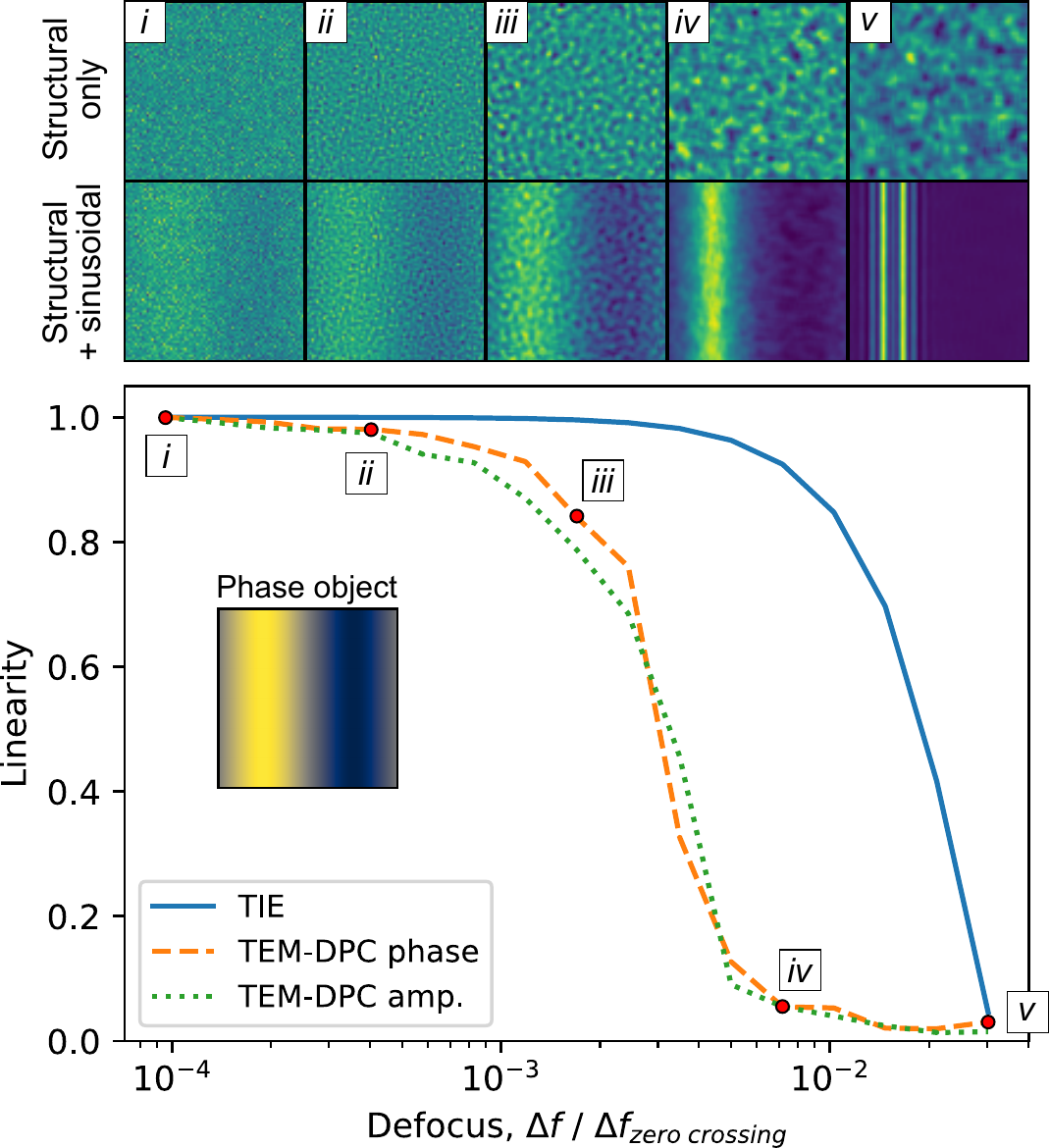}
      \caption{Example linearity of the TIE and TEM-DPC methods, assessed using an amplitude 10~radian sinusoidal phase object of period 5.12~$\muup$m (shown in the inset) at different defocus values.
      The defocus is normalised to the first CTF zero crossing for the spatial frequency of the sample phase.
      The two TEM-DPC profiles have white noise added to the sample to allow for alignment, either as pure phase or pure amplitude variations, as indicated in the legend.
      In both cases, the noise is filtered to present texture of an appropriate scale for non-rigid alignment.
      Example Fresnel images for phase structural contrast are shown in the top row, with the equivalent points marked as \textit{i - v} of the main panel.
      The TIE data was evaluated \textit{via} the gradient, $\partial I/\partial z$, calculated numerically with 0.001~m $z$-deltas, whereas the TEM-DPC data was obtained from alignment to images formed with only the structural phase.}
      \label{fig:linearity}
\end{figure}

To elucidate the relevant features, advantages and limitations of the TEM-DPC method and to compare it against the TIE one, we perform standard numerical image calculations, incorporating parallel illumination from a 200~kV source, defocus, and sample induced phase changes~\citep{DEGRAEF200127} for two illustrative cases: a sinusoidal phase object and a simple model of a magnetic domain wall.
We use the former to demonstrate linearity of the two methods and how intensity changes at large defocus limit the TEM-DPC method, while the latter is used to show how the dominant signal moves from intensity to deflection with increasing structural contrast.

\subsection{Linearity}
The main panel of Figure~\ref{fig:linearity} shows the linearity curves when using a sinusoidal pure phase object to provide the non-structural signal.
While the phase object is synthetic, we note that this situation is realised in the chiral helimagnet CrNb$_3$S$_6$~\citep{Togawa_PRL_2012_soliton}.
To provide a structural signal for image alignment, white noise is added to the sample for the two TEM-DPC profiles, either as pure phase (with amplitude 1) variations, or pure amplitude (with phase 0) variations.
As the measure of linearity, we use the ratio of the Fourier amplitude of the recovered phase component to that of the original non-structural phase.
The $x$-axis is the ratio of the defocus value to that of the first zero crossing of the CTF for the fixed frequency of the sinusoidal phase object.
Linearity is indicated by a $y$-axis value of 1, and is lost in both methods long before contrast reversal ($x$-axis values greater than 1).

The two TEM-DPC methods (dashed lines) have very similar linearity profiles; the source of the structural contrast is unimportant.
The exact profiles of both methods varies with the simulation parameters and, in this particular case, the TEM-DPC methods lose linearity at ${\sim }\,6\times$ lower defocus than does the TIE method (solid line).
This occurs because two factors adversely affect the image alignment.
First, the second derivative of the sinusoidal phase modifies the structural texture beyond what can be aligned to; and secondly, significant intensity modulations (precisely that which is used for the TIE analysis) become large.
These effects can be seen in the example Fresnel images with and without the sinusoidal phase that are shown in the top row of Fig.~\ref{fig:linearity} for different defocus values.
The latter physics can in principle be incorporated into the alignment algorithm, potentially extending the range of linearity of the TEM-DPC method.
However, as they stand, the results usefully demonstrate two general properties of the TEM-DPC method.
The first is that it is more suited to weakly varying phase gradients than is the TIE method; serendipitously, this is exactly where small intrinsic noise from structural contrast is typically present in many samples.
The second is somewhat related to the first and is that the TEM-DPC method is particularly well suited to samples where the structural contrast signal is significant compared to the non-structural signal.
We expand on this point in our second example, discussed next.

\subsection{Signal}
\begin{figure*}
  \centering
      \includegraphics[width=16.0cm]{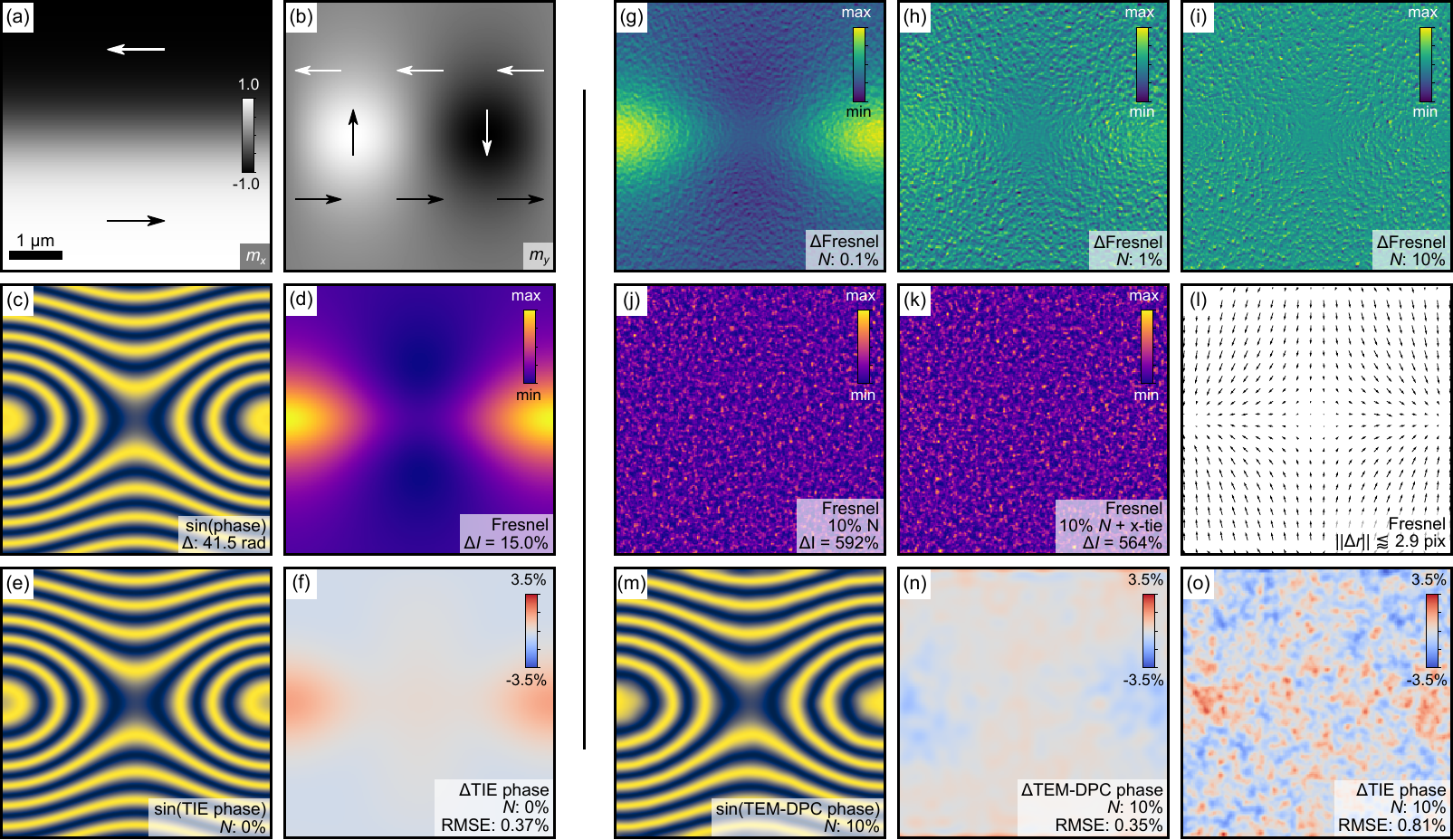}
      \caption{Cross-tie DW image simulations and analysis.
      (a), (b) Reduced magnetisation components of the DW and (c) the phase change determined from $\bm{m}$ through the Fourier-space approach~\citep{Beleggia_apl_2003_mag_sim}.
      (d) Fresnel image at $\Delta f$ = 8~mm (the same defocus was used for all data shown).
      (e) Sine of the TIE recovered phase and (f) the error in the phase.
      (g-i) Differential Fresnel images for different phase noise amplitudes, $N$, (0.1~\%, 1~\%, 10~\% of the magnetic phase range), representing structural contrast.
      Each panel is formed by subtracting from each Fresnel image one made with the same structural phase amplitude but with no magnetic phase.
      The noise was white, filtered with a Gaussian kernel with a 1~pixel standard deviation (image size: 256$\times$256 pixels).
      The Fresnel images for (i) are shown in (j) with structural phase only and (k) with structural plus magnetic phase.
      (l) Distortion vector field extracted from (j) and (k).
      (m) Sine of the TEM-DPC extracted phase and (n) the error in the phase.
      (o) The error in the TIE extracted phase for the data in (i).
      Where possible, the same colour map scale (colour and magnitudes) has been used across the same type of data.}
      \label{fig:TEM_DPC_noise_dependence}
\end{figure*}

We now consider simulations of a cross-tie magnetic domain wall (DW) with different levels of structural contrast.
The exact profiles used are not critical to the arguments made so, for simplicity, we use an analytical form of a cross-tie domain wall~\citep{Metlov_APL_2001_cross_tie}:
\begin{equation}
    \bm{m} = \{-\sinh(y/\lambda_c), \sin(x/\lambda_c), \cos(x/\lambda_c)\} / \cosh(y/\lambda_c),
    \label{eqn:cross_tie}
\end{equation}
where $\lambda_c$ is a constant proportional to the exchange length that controls the scaling.
This model has smoother features than is typical of cross-tie DWs, and was chosen as it resembles aspects of the experimental data reported in Part II~\citep{tem_dpc_part2}, which has a combination of uniform, circulatory and divergent magnetisation; the scaling here was also chosen to be somewhat similar to the experimental data.
Figure~\ref{fig:TEM_DPC_noise_dependence} summarises the simulation results.
To the left of the figure divide [Figs.~\ref{fig:TEM_DPC_noise_dependence}(a)--\ref{fig:TEM_DPC_noise_dependence}(f)], the system is defined and the results of analysis with no structural contrast are given.
The right hand side [Figs.~\ref{fig:TEM_DPC_noise_dependence}(g)--\ref{fig:TEM_DPC_noise_dependence}(o)] shows analysis results of the same system with structural contrast present.
In all panels, the same types of data are presented with the same colours to ease comparisons.

The phase change produced by the DW magnetisation [Figs.~\ref{fig:TEM_DPC_noise_dependence}(a) and \ref{fig:TEM_DPC_noise_dependence}(b)] is shown in Fig.~\ref{fig:TEM_DPC_noise_dependence}(c), and is periodic along the $x$-axis.
Throughout the figure, we plot the sine of the phase to highlight its contours.
The Fresnel image obtained at $\Delta f$= 8~mm [Fig.~\ref{fig:TEM_DPC_noise_dependence}(d)] shows bright regions at the left and right edges where a vortex is located, is saddle shaped in the central region where there is an anti-vortex, and has an intensity variation of 15~\% of the in-focus image intensity (denoted as $\Delta I$ in the figure annotations).
The phase recovered from this data using the TIE method [Fig.~\ref{fig:TEM_DPC_noise_dependence}(e)] is very similar to the true phase [Fig.~\ref{fig:TEM_DPC_noise_dependence}(c)], with small errors ($\sim$0.7~radians peak) at the vortices [Fig.~\ref{fig:TEM_DPC_noise_dependence}(f)], reflecting that the defocus is just outside the the linear regime for the sample (\emph{i.e.} most but not all of the phase is recovered).
The root mean square error (RMSE) is 0.37~\% of the true phase range.

For the remainder of the data in Fig.~\ref{fig:TEM_DPC_noise_dependence}, we add a white noise phase signal to represent structural contrast, after application of a Gaussian filter to reduce very high frequency components which are unrealistic (the effect of this is somewhat similar to that of an envelope function of the CTF).
The noise level is indicated by $N$, the amplitude as a percentage of the magnetic phase range.
The top row [Fig.~\ref{fig:TEM_DPC_noise_dependence}(g)--\ref{fig:TEM_DPC_noise_dependence}(i)] shows the difference between Fresnel images with the magnetic plus structural phase and the same with only identical structural phase, all at the same defocus, but with different structural phase amplitudes.
The displayed intensity range in each of these panels is matched to the data.
The first panel [Fig.~\ref{fig:TEM_DPC_noise_dependence}(g)] has a structural phase amplitude of 0.1~\% of the magnetic phase range, and appears generally similar to the Fresnel image with only magnetic contrast [Fig.~\ref{fig:TEM_DPC_noise_dependence}(d)], as one would expect.
The structural phase amplitude increases ten-fold in each following panel, reaching 10~\% of the DW phase range in Fig.~\ref{fig:TEM_DPC_noise_dependence}(i), where the structural contrast completely dominates the magnetism induced image contrast.
This occurs while the structural phase amplitude is only a small percentage of the magnetic phase range because it is the second order phase gradients that determine the intensity variations and these are much higher for high frequency signals.

The two Fresnel images that form the difference image with the highest structural contrast [Fig.~\ref{fig:TEM_DPC_noise_dependence}(i)] are shown in Figs.~\ref{fig:TEM_DPC_noise_dependence}(j) and \ref{fig:TEM_DPC_noise_dependence}(k), and although none of the images show any obvious sign of magnetic contrast from the \emph{intensity} contrast, there is a magnetic signal in the \emph{structural} contrast.
The displacement field extracted from this data [Fig.~\ref{fig:TEM_DPC_noise_dependence}(l)] shows a very clear pattern from the magnetic signal.
The maximum magnitude of the vector field, $\bm{\Delta r}$, is small ($\lessapprox$~2.9~pixels), and so it would be very easy to dismiss such a sample as non-magnetic by visual examination of the Fresnel images alone.
The locations of the vortex and anti-vortex can be readily identified by the vector field magnitude reducing to zero, and differentiated by the orientation of the arrows.
The locations of the vortex and anti-vortex cores are also visible in Fig.~\ref{fig:TEM_DPC_noise_dependence}(i) as regions with reduced contrast.
This occurs because there is no image displacement at the core centres, and thus the noise signal increasingly cancels towards those locations.

The phase change from the magnetic induction [Fig.~\ref{fig:TEM_DPC_noise_dependence}(m)] obtained from the extracted displacements using Eq.~(\ref{eqn:phasegrad_displacement}) also matches the true phase [Fig.~\ref{fig:TEM_DPC_noise_dependence}(c)] very well.
The error in the phase [Fig.~\ref{fig:TEM_DPC_noise_dependence}(n)] is somewhat less peaked than the TIE error [Fig.~\ref{fig:TEM_DPC_noise_dependence}(f)], with a slight increase at the top and bottom edges due to missing overlap of between the images as a result of the image distortion.
This missing region does not appear as significant in the TIE analysis only because the phase varies relatively slowly at those locations, and so the intensities at the different locations are similar; this will not be the case elsewhere, such as at sample edges.
Including these edges, the RMSE of the extracted phase is 0.35~\%, slightly smaller than that of the TIE method (excluding the edges gives a value of 0.28~\%).
The error in the TIE extracted phase from the same images [Fig.~\ref{fig:TEM_DPC_noise_dependence}(o)] has a similar overall shape to that from the images with no structural contrast [Fig.~\ref{fig:TEM_DPC_noise_dependence}(f)], as one might expect.
The additional speckle signal is from the structural contrast and this can be partially removed by filtering the source images with the same Gaussian filter ($\sigma = 8$ pixels) used in the TEM-DPC alignment regularisation, resulting in an RMSE of 0.56~\%.

Beyond demonstrating that the TEM-DPC technique works well in analysing data with very large structural contrast, the simulations above also make clear that the greatest signal in the data moves from intensity to deflection at higher levels of structural contrast.
For the 10~\% structural phase amplitude data examined above, the intensity range in the Fresnel images [Figs.~\ref{fig:TEM_DPC_noise_dependence}(j) and \ref{fig:TEM_DPC_noise_dependence}(k)] is approximately 40$\times$ higher than that in the Fresnel image obtained from the purely magnetic sample [Fig.~\ref{fig:TEM_DPC_noise_dependence}(d)], corresponding to a $\sim$6-fold increase in peak intensity.
In this situation, high dynamic range detectors and long exposures may be needed in experiment to avoid small electron counts causing reduction in the signal-to-noise ratio of the TIE signal.
However, even with high dynamic range imaging, the TEM-DPC method may be much more efficient and practical to use for this type of data.

\subsection{Discussion}
The exact conditions under which use of the \mbox{TEM-DPC} method will produce more accurate results than the TIE technique depends on several factors: (1) the amplitude and spectral components of the structural contrast; (2) the amplitude and spectral components of the non-structural (magnetic or otherwise) contrast; (3) the magnitude of the defocus used; and, as always, (4) the quality of the optical alignment, including in particular other aberrations and changes in illumination conditions that come with defocus.
The examples above illustrate aspects of points (1) -- (3).
Point (4) has a greater bearing on practical matters.

Even in well aligned microscopes, it is inevitable that there will be small image shifts or sample position changes between successive images at different defocus levels which will require the images to be rigidly aligned.
Importantly, there may also be changes in magnification, in illumination brightness, centering or uniformity, and in optical distortions, particularly at larger defocus values, potentially giving rise to systematic errors.
These will all adversely affect the TIE method, but since the TEM-DPC method uses positional rather than intensity changes, and in the version of the method employed here, uses as a reference an image obtained at the same defocus, it is largely unaffected by these factors (the experimental data of Part II~\citep{tem_dpc_part2} provides an example of this).
Of course, a method to alter the polarisation of the sample is required for this version of the TEM-DPC method, but this is often already an integral part of many \textit{in-situ} experiments.

\section{Conclusions}
We have outlined a methodology for quantitative phase recovery from Fresnel imaging in a standard TEM that has origins dating back half a century, but that has since gone largely unused in electron microscopy.
The technique, which we refer to as TEM-DPC, uses image distortion rather than intensity changes and can be applied to any source of phase contrast in an unmodified TEM, provided a reference image with suitable contrast and of a known state can be obtained.

Through numerical simulations we have demonstrated the main features of the technique, and discussed some of the practical advantages it has over the commonly used TIE method.
In particular, the TEM-DPC method can be an invaluable tool to analyse samples with relatively strong structural contrast compared with phase contrast.
This applies both to samples with high levels of structural contrast, where there may be signal-to-noise ratio issues in the TIE method, and to samples with low phase contrast, where ever higher levels of microscope alignment and stability are required to avoid systematic errors affecting that method.
Under which exact circumstances each technique will provide more benefit than the other will depend on specific details of the sample and the experiment.
Where one method does not have an obvious advantage over the other in a particular application, then a hybrid approach, where aspects of the two models are combined in order to maximise use of all information contained in the data, may prove beneficial.
However, in general, compared with the TIE method, the TEM-DPC method is intrinsically less susceptible to common systematic errors of magnification, distortion, rotation, and illumination changes that are often encountered when changing focus.

We expect the TEM-DPC method to be of potential use in the study of a broad range of samples, especially as \textit{in-situ} experiments are becoming more common due to the additional insight they yield.
Such experiments often produce suitable reference images, allowing the technique presented to be applied to the same dataset as would be collected for TIE analysis, with little or no requirement for additional measurements.
We demonstrate this and other aspects of the technique using experimental data in Part II of this work, which also includes details of the magnetic phase transition in K$_2$CuF$_4$~\citep{tem_dpc_part2}.

Original data files for the work reported herein are available at DOI: TBA.
All substantive simulation, analysis, and visualisation code is freely available in the open source \texttt{fpd} Python library~\citep{fpd}.

\noindent\small\color{Maroon}\textbf{Acknowledgements }\color{Black}
We acknowledge support from the Engineering and Physical Sciences Research Council (EPSRC) of the United Kingdom (Grant Number EP/M024423/1); Grants-in-Aid for Scientific Research on Innovative Areas `Quantum Liquid Crystals' (KAKENHI Grant No. JP19H05826) from JSPS of Japan; Grants-in-Aid for Scientific Research (KAKENHI grant Nos. 17H02767 and 17H02923) from JSPS of Japan; and the Carnegie Trust for the Universities of Scotland.

% \renewcommand{\thefigure}{A\arabic{figure}}
% \setcounter{figure}{0}  
% 
% \appendix

\normalsize
% authors generating their own bbl file would uncomment the following two lines, and comment out/delete the references below:

%\bibliography{igsrefs}   % reads igsrefs.bib
%\bibliographystyle{igs}  % imposes IGS bibliography style on output
\bibliographystyle{MandM}
% \bibliography{refs.bib}

% \onecolumngrid
\balance

\end{document}